\documentclass[preprint,showpacs,preprintnumbers,amsmath,amssymb,superscriptaddress,bibnotes, apl]{revtex4-1}

\usepackage{graphicx}% Include figure files
\usepackage{dcolumn}% Align table columns on decimal point
\usepackage{tabularx}
\usepackage{bm}% bold math
\usepackage{color}
\makeatletter
\def\Ddots{\mathinner{\mkern1mu\raise\p@	
\vbox{\kern7\p@\hbox{.}}\mkern2mu
\raise4\p@\hbox{.}\mkern2mu\raise7\p@\hbox{.}\mkern1mu}}
\makeatother

\begin{document}

\preprint{Preprint}

\title{A nanoscale vacuum-tube diode triggered by few-cycle laser pulses}% Force line breaks with \\

\author{Takuya Higuchi} 
\email[E-mail: ]{takuya.higuchi@fau.de}
\affiliation{Department of Physics, Friedrich-Alexander-Universit\"at Erlangen-N\"urnberg, Staudtstrasse 1, D-91058 Erlangen, Germany}
\affiliation{Max-Planck-Institut f\"ur Quantenoptik, Hans-Kopfermann-Strasse 1, D-85748 Garching, Germany}\author{Lothar Maisenbacher}
\affiliation{Max-Planck-Institut f\"ur Quantenoptik, Hans-Kopfermann-Strasse 1, D-85748 Garching, Germany}
\author{Andreas Liehl}
\email[Current address: ]{Department of Physics, University of Konstanz, D-78457 Konstanz, Germany}
\affiliation{Max-Planck-Institut f\"ur Quantenoptik, Hans-Kopfermann-Strasse 1, D-85748 Garching, Germany}
\author{P\'eter Dombi}
\email[Current address: ]{MTA ``Lend\"ulet'' Ultrafast Nanooptics Group, Wigner Research Centre for Physics 1121 Budapest, Hungary}
\affiliation{Max-Planck-Institut f\"ur Quantenoptik, Hans-Kopfermann-Strasse 1, D-85748 Garching, Germany}
\author{Peter Hommelhoff}
\email[E-mail: ]{peter.hommelhoff@physik.uni-erlangen.de}
\affiliation{Department of Physics, Friedrich-Alexander-Universit\"at Erlangen-N\"urnberg, Staudtstrasse 1, D-91058 Erlangen, Germany}
\affiliation{Max-Planck-Institut f\"ur Quantenoptik, Hans-Kopfermann-Strasse 1, D-85748 Garching, Germany}
\date{\today}% It is always \today, today,

\begin{abstract}
We propose and demonstrate a nanoscale vacuum-tube diode triggered by few-cycle near-infrared laser pulses. It represents an ultrafast electronic device based on light fields, exploiting near-field optical enhancement at surfaces of two metal nanotips. The sharper of the two tips displays a stronger field-enhancement, resulting in larger photoemission yields at its surface. One laser pulse with a peak intensity of $4.7\times 10^{11}$ W/cm$^2$ triggers photoemission of ${\sim }16$ electrons from the sharper cathode tip, while emission from the blunter anode tip is suppressed by $19$ dB to ${\sim }0.2$ electrons per pulse. Thus, the laser-triggered current between two tips exhibit a rectifying behavior, in analogy to classical vacuum-tube diodes. According to the kinetic energy of the emitted electrons and the distance between tips, the total operation time of this laser-triggered nanoscale diode is estimated to be below 1 ps.
\end{abstract}
%\pacs{72.20.Ht, 42.65.Ky, 42.65.Re}% PACS, the Physics and Astronomy

%72.20.Ht	High-field and nonlinear effects
%42.65.Ky	Frequency conversion; harmonic generation, including higher-order harmonic generation 
%42.65.Re	Ultrafast processes; optical pulse generation and pulse compression

\maketitle

Ultrashort-pulsed lasers are nowadays widely employed to steer electrons on time scales that are orders of magnitude shorter than those of the existing electronics devices \cite{Krausz:2014dn,Schiffrin:2013rc,Ghimire:2011fk, Ivanov20133}. For example, photoemission by femtosecond pulsed lasers generates free electrons that are confined within the laser pulse durations. These ultrashort electron pulses have enabled various ultrafast measurements such as time-resolved electron diffraction \cite{Siwick21112003,Miller07032014,Baum02112007} and microscopy \cite{Yang24082010}.

It is tempting to utilize these pulsed electrons as carriers in electronic devices because the capability to trigger them by laser pulses bares the potential to overcome the limitations of the operation speed of present electronic devices. The concept of using {\it free} electrons as carriers resembles classical vacuum-tube devices, in which electrons emitted via thermionic emission are employed as carriers \cite{PrinciplesOfElectronTubes}. The functions of the tube devices are achieved via control of the potential of electrodes that determine the trajectories of these electrons, which is readily applicable to control those of the laser triggered electrons as well.

However, replacing the thermal cathode by laser-triggered emission alone is not sufficient to achieve ultrafast electronic devices. This is because the conventional tube electronics exhibit macroscopic (millimeters to centimeters) length scale, and the operation speed of the device is mainly determined by the traveling time of the electrons from one electrode to the other \cite{PrinciplesOfElectronTubes}. To take advantage of prompt laser-triggered photoemission, a reduction of the electron travel distance is necessary, which leads us to downsize the vacuum-tube devices drastically. Namely, a photoelectron with a typical kinetic energy of $1$ eV has a velocity of $0.6$ ${\mu }$m/ps in vacuum, and thus the channel distance should be within sub-micrometer length scales to achieve triggering operation faster than 1 THz (i.e., $(1$ ps$)^{-1}$). 

In this study, we propose a geometry of electronic devices that consist of metal nanotips illuminated by laser pulses. As a first example, we demonstrate a nanoscale vacuum-tube diode consisting of two metal nanotips, which work as cathode (emitting electrons) and anode (capturing electrons). 

A key aspect of the metal nanotips is found in the enhancement of the optical near fields at their apexes. These near fields are confined to length scales much smaller than the wavelength of light \cite{Novotny-book}. They are widely employed to improve spatial resolution in the scanning near-field optical microscopy \cite{:/content/aip/journal/jcp/112/18/10.1063/1.481382} and the tip-enhanced Raman spectroscopy \cite{Stockle:2000rp}. This localized field enhancement also plays an important role in electron emission, resulting in generation of spatiotemporally confined electron pulses \cite{PhysRevLett.105.257601,Kruger:2011kx,PhysRevLett.103.257603,Herink:2012pi,Wimmer:2014rw, Hoffrogge2014}. 

One can taylor the strength of the optical near fields by controlling the tip radius: the smaller the tip radius, the larger the field enhancement factor \cite{Novotny-book,doi:10.1021/nl402407r}. Due to the stronger field enhancement at the sharper tip, the electron emission yield is much larger there than at the other tip, even if both tips are within the same laser spot. As a result, electrons are mainly emitted from the sharper tip under laser excitation, and this tip consequently works as a cathode. On the other hand, the blunter tip barely emits electrons and receives the electrons emitted from the cathode tip; thus it works as an anode.
In this study, we use a comparably blunt tip as anode to take advantage of the well-characterized surface of the tip, as well as to utilize the superior optical access provided by the shape of the tip. 
Also, the small absorption cross section of a nanotip is advantageous compared with a flat surface.

\begin{figure}[t]
\begin{center}
\includegraphics[width=8.5cm]{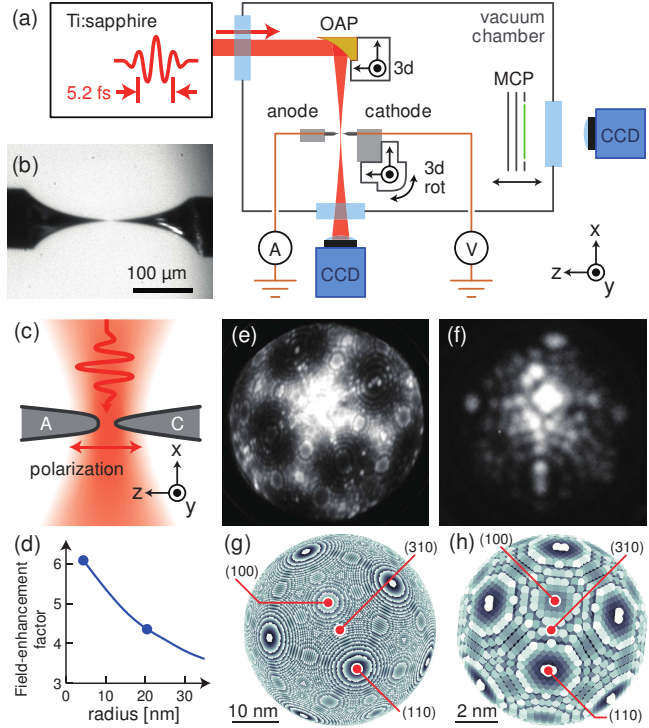}% Here is how to import EPS art
\caption{\label{Figure1} (a) Schematics of the experimental set up. OAP: off-axis parabolic mirror, CCD: charge-coupled device camera, MCP: microchannel plate electron detector. (b) Optical microscopic image of the two metal nanotips facing each other. (c) Schematics of the two tips under illumination of a few-cycle-laser pulses, showing both two tips within the laser spot. A and C mean anode and cathode, respectively. 
(d) Simulated values of the maximum field enhancement factor at the surface of a tungsten tip as a function of tip radius. Data taken from \cite{doi:10.1021/nl402407r}. The two points indicate the radii of the two tips used in this study. Field ion microscope images of (e) the blunter anode tip and (f) the sharper cathode tip. The corresponding protrusion maps (light:~protrusion, dark:~depression) on the basis of a ball model for bcc structures are depicted in (g) and (h). From theses images we infer tip radii of $20.4\pm 0.8$ for the anode tip and $4.4\pm 0.8$ nm for the cathode tip.
}
\end{center}
\end{figure}

Figure~\ref{Figure1}(a) shows the schematics of the experimental setup. 
The cathode and the anode are two tungsten tips that are produced by electro-chemical etching from a single-crystalline tungsten wire in [310] orientation. 
The tips are placed in an ultrahigh vacuum chamber with a base pressure of $10^{-7}$ Pa.
The cathode tip is mounted on a 3d piezo-electric translational stage and faces the fixed anode tip, as shown in the optical microscopic image in Fig.~\ref{Figure1}(b).
After the tips have been approached (described later), they are illuminated by the laser as schematically depicted in Fig.~\ref{Figure1}(c).
We use two-optical-cycle laser pulses from a Ti:Sapphire laser oscillator with a repetition rate of 80 MHz and a center wavelength of 780 nm. The Fourier-transform-limited pulse duration is 5.2 fs (full width at half maximum of the intensity envelope). The dispersion is compensated so that the shortest pulse duration is achieved at the position of the tips, as determined from an interferometric autocorrelation trace. The laser beam is focused on the tips by an off-axis parabolic mirror (OAP), which is mounted on another piezo-electric stage. The laser spot position can be moved by shifting the position of the OAP.
The laser spot size has a $1/e^2$ radius of $2.3$ $\mu$m. These laser parameters result in a peak intensity of $5.6\times10^{11}~{\rm W/cm^2}$ and a peak electric field of $2.1~{\rm V/nm}$ at the bare focus in free space for an average power of $20~{\rm mW}$. The electric field is enhanced around the apex of each tip depending on its radius of curvature, as shown in the simulation results in Fig.~\ref{Figure1}(d), which are supported by experimental data \cite{doi:10.1021/nl402407r}. Also, when the two tips are apart further than the tip radii, the near-field enhancement at one tip is not influenced by the presence of the other tip. This condition is always satisfied in this paper.

To measure and control the radii of the two tips, in-situ field ion microscopy (FIM) and evaporation are available \cite{PhysRev.102.624}.
For this purpose, both tips can be directed to a microchannel-plate electron detector (MCP) independently, which is achieved by placing a rotational stage underneath the 3d translational stage for the cathode tip. FIM with He as an image gas enables us to observe the surface structures of the tips with atomic resolution [Figs.~\ref{Figure1}(e) and (f). See also the ball-model structures in Figs.~\ref{Figure1}(g) and (h)]. During the FIM process, a positive voltage (typically 5 kV $\sim$ 15 kV) is applied to obtain an image of the tip apex with the help of field-ionized He ions. By applying a voltage higher ($+{\sim }0.5$ kV) than the best imaging voltage, the tungsten atoms are field-evaporated. Keeping a tip under this condition, it can be blunted in a controlled manner. An as-grown tip typically has a radius of curvature of $5\sim 10$ nm. In this study, we blunted the anode tip to a radius of curvature of $20.4\pm 0.8$ nm, which was determined by counting the number of rings between different crystalline facets \cite{Tsong-book}. The cathode tip is much sharper, and has a radius of curvature of $4.4\pm 0.8$ nm.

\begin{figure}[t]
\begin{center}
\includegraphics[width=8.5cm]{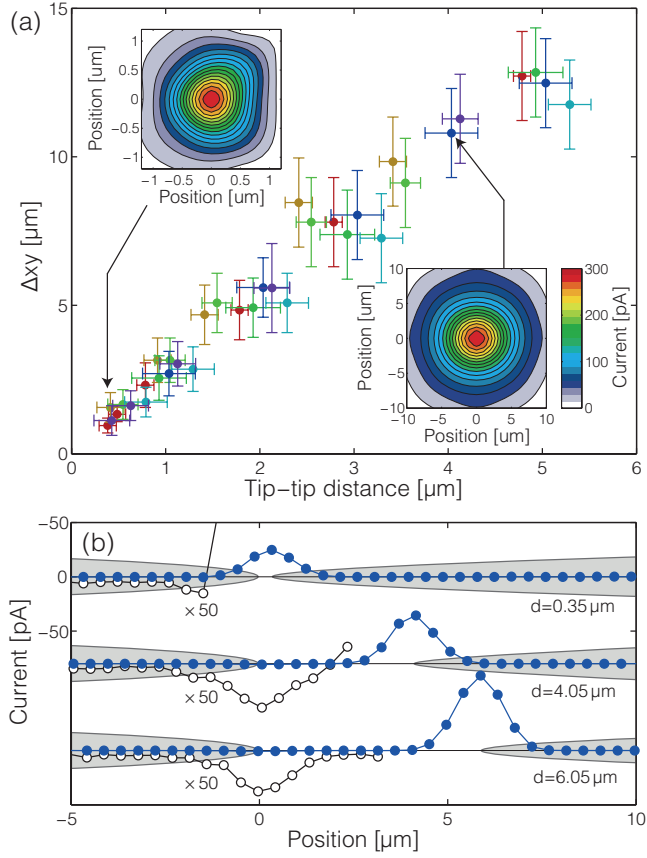}% Here is how to import EPS art
\caption{\label{Figure2} 
Approaching of the two tips:
(a) Full width at half maximum ($\Delta xy$) of the field emission current mapping as a function of the tip-tip distance.
Insets show the field emission current mapping at two different tip-tip distance, $d = 0.35$ ${\rm \mu m}$ and $d = 4.05$ ${\rm \mu m}$. The tip approach is carried out several times, and data points with one color are one set of data taken during one approach trial.
(b) Laser induced current between two tips under illumination of the laser, and as a function of the laser spot position. Curves for three different tip distances are shown. The plots are offset from each other to enhance readability. Magnified (factor-of-50) plots are also shown to make the peak around the anode tip visible.
When the sign of the current is positive, the electrons travel from the cathode to the anode.
The shadows schematically depict the position of the two tips. The fixed tip (left) is the anode.
}
\end{center}
\end{figure}

We align the two tips on the same axis and bring them to a distance of $d = 350 \pm 100$ nm with the help of the DC-field emission current between the two tips, which depends sensitively on the relative position of the tips.
When we scan the position of the cathode tip in a plane perpendicular to the tip axis, the current depends on this position as shown in the insets of Fig.~\ref{Figure2}(a). We tune the voltage (typically cathode tip at ${\sim }-200$ V at a distance of 1 $\mu$m) so that the maximum current in the 2D map is ${\sim }300$ pA. The data are well fitted assuming that the field strength is inversely proportional to the distance ($=\sqrt{x^2+y^2+z^2}$) between them, and that the tunneling current follows the Fowler-Nordheim behavior \cite{Fowler:1928}. The tips can be aligned on the tip-tip axis by laterally finding the peak in the current. 

After the lateral tip adjustment, a better estimation of the tip-tip distance is made by fitting the width $\Delta xy$ of the lateral ($xy$) distributions [see Fig.\ref{Figure2}(a)] as a function of the $z$-position. Although the absolute distance between the tips cannot be directly measured without crashing the two tips, one can approach one tip to the other step-by-step, and the distances between the steps can be determined with an accuracy of $60$ nm by the interferometric scale built inside the translational stage. We find a linear scaling between $\Delta xy$ and the tip distance by this step-by-step approach, and extrapolated the position $d=z=0$ as the position $\Delta xy$ becomes zero. Figure~\ref{Figure2}(a) shows $\Delta xy$ as a function of the tip distance. Here, the data sets during different approach routines are plotted in the same figure, distinguished by the colors of the data points. For all the data set, the distance between the two tips can be derived with a resolution of $100$ nm, without crashing the two tips. Note that the resolution is now limited by the mechanical stability of the experimental system, as we proved experimentally.

Now we focus the laser pulses on the two tips. Figure \ref{Figure2}(b) shows the measured current between the laser-triggered cathode and the anode while the laser spot is scanned along their axes. No bias voltage is applied. The two tips are pre-aligned onto a same axis, and for three different tip-tip distances $d$, the laser-position dependence of the current is measured.
When the two tips are apart further than the laser spot radii, we measure two peaks with opposite signs in the current as a function of the beam position, corresponding to the emission from the anode and the cathode (bottom in Fig.~\ref{Figure2}(b)). The emission from the cathode tip
is much larger than that of the anode because of the following reasons:
Electrons are emitted by absorption of at least three photons of energy $\hbar \omega \approx 1.58$ eV
as the work function $\phi$ of tungsten is 4.35 eV at the W(310) facet \cite{Kawano:2008rm}.
The dominating part of photocurrent is due to three- and four-photon photoemission, without a bias voltage.
Hence, the effect of the larger field enhancement factor of the sharper tip is accentuated by the nonlinear nature of the emission process.
This explains why the emission current from the sharper cathode tip is ${\sim }50$ times larger than that of the blunter anode tip. 

\begin{figure}[t]
\begin{center}
\includegraphics[width=8.5cm]{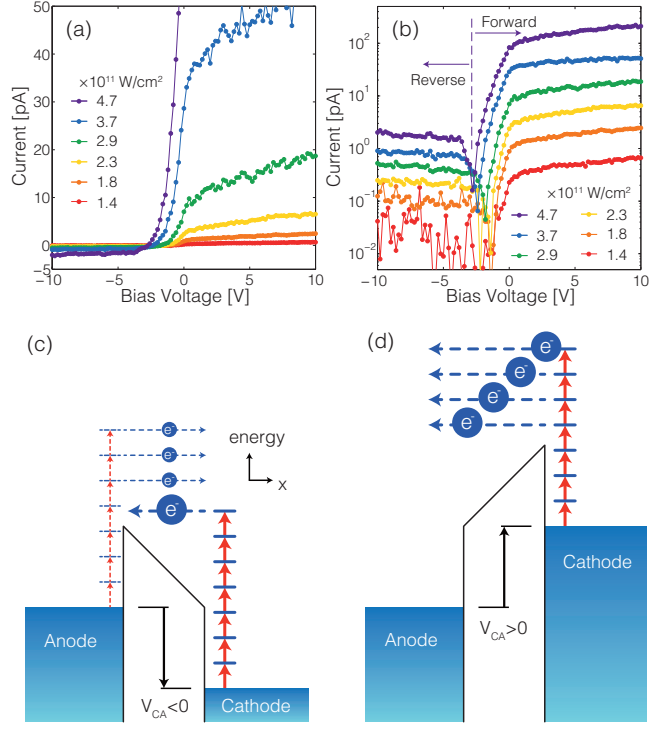}% Here is how to import EPS art
\caption{\label{Figure3} 
Operation of the laser-triggered nanoscale vacuum-tube diode: 
(a) Linear-scale plot of current as a function of the bias voltage. Various curves are recorded for different laser peak intensities. (b) Log-scale plot of (a). Here, the absolute values of the current are plotted (signs change close to the minima). As an example, the dashed line indicates the point of change in sign for the strongest laser intensity. Schematic diagrams of the device operation under (c) reverse bias voltage condition and (d) forward bias voltage condition.}
\end{center}
\end{figure}

By approaching the two tips ($d = 350 \pm 100$ nm, topmost plot in Fig.~\ref{Figure2}(b)) within the beam spot (radius 2.3 $\mu$m), the weaker emission from the anode is submerged by the stronger emission from the cathode. A pair of tips under this condition is expected to work as a diode because electrons should be mainly emitted from the cathode.

Figure~\ref{Figure3} shows the operation of this laser-triggered diode.
Figures~\ref{Figure3}(a) and (b) show the measured current between the laser-triggered cathode and the anode tips as a function of the bias voltage $V_{\rm CA}$ between them.
It depends nonlinearly on the bias voltage and shows a rectifying behavior.
The forward saturation current is 210 pA for the largest laser intensity ($4.7 \times 10^{11}$ W/cm$^2$ nominal laser intensity in the bare focus, i.e. without field enhancement). The repetition rate of the laser is 80 MHz, and thus this corresponds to ${\sim }16$ electrons per pulse on average. On the other hand, the reverse saturation current is 2.0 pA, which is two orders of magnitude smaller than the forward one. 

The mechanism of diode operation is explained using the energy diagrams depicted in Figs.~\ref{Figure3}(c) and (d).
Electrons are emitted through multi-photon photoemission. 
When a reverse bias is applied [Fig.~\ref{Figure3}(c)], 
the current is reduced because only electrons having higher energies than the potential barrier can reach the anode. 
The weak current in strong reverse bias conditions results from the weak emission from the anode tip.
On the other hand, a forward bias allows all the electrons that overcome the work function to flow [Fig.~\ref{Figure3}(d)]. 
The short-circuit current increases as the laser intensity increases.

When a reverse bias is applied, only electrons with higher energy than the bias potential can reach the anode.
This determines the lowest initial velocity and thus the longest traveling time $\tau_L$ of electrons that contribute to the current. Assuming a plate-capacitor potential, a tip-tip distance of $d=350$ nm and $V_{\rm CA} = -2$ V, a conservative estimate yields $\tau_L = 800$ fs. 
The photoemission process takes place on the time scale of the laser pulse duration ($5.2$ fs), thus the whole process including the electron emission and current flow between two electrodes completes on sub-picosecond timescales. In the forward-biased condition, one can reduce the traveling time below $\tau_L$ by acceleration due to the forward field, which is not limited by the initial kinetic energy unlike the case of reverse-biased condition.

Note that electrons are directionally emitted along the axis of the tip (i.e., the [310] crystalline axis in this work) because the (310) facet of tungsten has the lowest work function \cite{PhysRevLett.103.257603}.
With a typical opening angle of the (310) electron beam of $\sim$ 10 degrees, the spread in longitudinal velocity caused by this small angular distribution is $\sim$1 percent, which is not significant.
We also numerically estimated the effect of Coulomb repulsion by using a charged-particle-dynamics simulator (General Particle Tracer) as one electron bunch contains $\sim$16 electrons per laser pulse at maximum. For this estimation, the key factor is the initial volume of the bunch. Namely, the emission area is restricted by the radius of cathode tip ($r=4.4$ nm), and emission is confined temporally within the laser pulse duration, leading to an initial bunch length of $<10$ nm.
For such a small initial volume, the Coulomb repulsion between electrons occurs significantly only within $\sim$10 fs after emission, expanding the electron cloud to $\sim$ 25 nm in radius. After this initial space-charge expansion, the electrons behave almost independently. Therefore, we can neglect corrections by the space-charge effect to the above value of $\tau_L=800$ fs as estimated on the basis of the final kinetic energy.

To summarize, we observed rectifying behavior between two metal nanotips, 
where the electrons are emitted from the cathode tip through multi-photon photoemission by few-cycle laser pulses. The photoemission properties of the two tips were controlled by tuning the tip-radii.
This diode device is switched by the laser pulse and operates on sub-picosecond timescales, owing to the short duration of the electron emission process, the comparably high initial kinetic energy of the emitted electrons,
and the sub-micron distance between the two electrodes.
Reducing the tip-tip distance will allow entering further important regimes for electrons and light fields to interact: when the tip-tip distance is smaller than the near-field decay length, i.e., the radii of curvature of the tips, the optical near fields around the tips are not independent any more and to exhibit stronger field enhancement. Moreover, when the tip-tip distance is within the decay length of the electron's wave functions, i.e., sub-nanometer length scales, quantum tunneling channels are formed \cite{Savage:2012dk}. These tunneling channels can be steered by the electric field of light, as expected from the observations that electron emission from nanotips depends on the carrier-envelope phase of few-cycle laser pulses \cite{Kruger:2011kx}. Together with these advanced methods, our achievement of diode operation with metal nanotips has the potential to operate on attosecond timescales, which could open a way to petahertz electronics operating at optical frequencies.

The authors acknowledge M.~Kr{\"u}ger and M.~F{\"o}rster for their technical support, as well as thank S.~Thomas for providing the field-enhancement simulation results.
This work was supported by the DFG Cluster of Excellence Munich-Centre for Advanced Photonics
and the ERC grant ``Near Field Atto.''
TH acknowledges a JSPS fellowship for research abroad.
PD acknowledges a Marie Curie Fellowship ("UPNEX", project number 302657).

%\bibliographystyle{prsty_noetal}
%\bibliography{../../Literature.bib}

\end{document}